\documentclass[10pt,conference,romanappendices]{IEEEtran}
\usepackage{fancyhdr} 
\usepackage[british,UKenglish,USenglish,american]{babel}
\usepackage{epsfig,graphics,subfigure,graphicx,latexsym,longtable,amsmath,amscd,latexsym,amssymb,mathrsfs,syntonly,eucal}
\usepackage{cite}
\usepackage{multirow}
\usepackage[usenames]{color}
\usepackage[T1]{fontenc}
\usepackage{bm,cite}
\usepackage{amsbsy}
\usepackage{latexsym}
\usepackage{wasysym}
\usepackage{placeins}
\usepackage[lined,boxed,commentsnumbered]{algorithm2e}
\usepackage{lipsum}
\usepackage{url}
\usepackage{array}
\usepackage{tabu}
\usepackage[skip=4pt,font=scriptsize]{caption}
\usepackage{soul}
\usepackage{geometry}
\SetKwInput{KwInput}{Input}
\SetKwInput{KwOutput}{Output}

\usepackage[acronyms,nonumberlist,nopostdot,nomain,nogroupskip,acronymlists={hidden}]{glossaries}
\newglossary[algh]{hidden}{acrh}{acnh}{Hidden Acronyms}
\glsdisablehyper
\newacronym{3gpp}{3GPP}{3rd Generation Partnership Project}
\newacronym{4g}{4G}{4th generation}
\newacronym{5g}{5G}{5th generation}
\newacronym{5gc}{5GC}{5G Core}
\newacronym{adc}{ADC}{Analog to Digital Converter}
\newacronym{aerpaw}{AERPAW}{Aerial Experimentation and Research Platform for Advanced Wireless}
\newacronym{ai}{AI}{Artificial Intelligence}
\newacronym{aimd}{AIMD}{Additive Increase Multiplicative Decrease}
\newacronym{am}{AM}{Acknowledged Mode}
\newacronym{amc}{AMC}{Adaptive Modulation and Coding}
\newacronym{amf}{AMF}{Access and Mobility Management Function}
\newacronym{aops}{AOPS}{Adaptive Order Prediction Scheduling}
\newacronym{api}{API}{Application Programming Interface}
\newacronym{apn}{APN}{Access Point Name}
\newacronym{aqm}{AQM}{Active Queue Management}
\newacronym{ausf}{AUSF}{Authentication Server Function}
\newacronym{avc}{AVC}{Advanced Video Coding}
\newacronym{awgn}{AGWN}{Additive White Gaussian Noise}
\newacronym{balia}{BALIA}{Balanced Link Adaptation Algorithm}
\newacronym{bbu}{BBU}{Base Band Unit}
\newacronym{bdp}{BDP}{Bandwidth-Delay Product}
\newacronym{ber}{BER}{Bit Error Rate}
\newacronym{bf}{BF}{Beamforming}
\newacronym{bler}{BLER}{Block Error Rate}
\newacronym{brr}{BRR}{Bayesian Ridge Regressor}
\newacronym{bsr}{BSR}{Buffer Status Report}
\newacronym{bs}{BS}{Base Station}
\newacronym{bss}{BSS}{Business Support System}
\newacronym{ca}{CA}{Carrier Aggregation}
\newacronym{caas}{CaaS}{Connectivity-as-a-Service}
\newacronym{cb}{CB}{Code Block}
\newacronym{cc}{CC}{Congestion Control}
\newacronym{ccid}{CCID}{Congestion Control ID}
\newacronym{cco}{CC}{Carrier Component}
\newacronym{cdd}{CDD}{Cyclic Delay Diversity}
\newacronym{cdf}{CDF}{Cumulative Distribution Function}
\newacronym{cdn}{CDN}{Content Distribution Network}
\newacronym{cir}{CIR}{Channel Impulse Response}
\newacronym{cn}{CN}{Core Network}
\newacronym{codel}{CoDel}{Controlled Delay Management}
\newacronym{comac}{COMAC}{Converged Multi-Access and Core}
\newacronym{cord}{CORD}{Central Office Re-architected as a Datacenter}
\newacronym{cornet}{CORNET}{COgnitive Radio NETwork}
\newacronym{cosmos}{COSMOS}{Cloud Enhanced Open Software Defined Mobile Wireless Testbed for City-Scale Deployment}
\newacronym{cots}{COTS}{Commercial Off-the-Shelf}
\newacronym{cp}{CP}{Control Plane}
\newacronym{cpu}{CPU}{Central Processing Unit}
\newacronym{cqi}{CQI}{Channel Quality Information}
\newacronym{cr}{CR}{Cognitive Radio}
\newacronym{cran}{CRAN}{Cloud \gls{ran}}
\newacronym{crs}{CRS}{Cell Reference Signal}
\newacronym{csi}{CSI}{Channel State Information}
\newacronym{csirs}{CSI-RS}{Channel State Information - Reference Signal}
\newacronym{cu}{CU}{Central Unit}
\newacronym{d2tcp}{D$^2$TCP}{Deadline-aware Data center TCP}
\newacronym{d3}{D$^3$}{Deadline-Driven Delivery}
\newacronym{dac}{DAC}{Digital to Analog Converter}
\newacronym{dag}{DAG}{Directed Acyclic Graph}
\newacronym{darpa}{DARPA}{Defense Advanced Research Projects Agency}
\newacronym{das}{DAS}{Distributed Antenna System}
\newacronym{dash}{DASH}{Dynamic Adaptive Streaming over HTTP}
\newacronym{dc}{DC}{Dual Connectivity}
\newacronym{dccp}{DCCP}{Datagram Congestion Control Protocol}
\newacronym{dce}{DCE}{Direct Code Execution}
\newacronym{dci}{DCI}{Downlink Control Information}
\newacronym{dcl}{DCL}{Dear Colleague Letter}
\newacronym{dctcp}{DCTCP}{Data Center TCP}
\newacronym{dl}{DL}{Downlink}
\newacronym{dmr}{DMR}{Deadline Miss Ratio}
\newacronym{dmrs}{DMRS}{DeModulation Reference Signal}
\newacronym{drlcc}{DRL-CC}{Deep Reinforcement Learning Congestion Control}
\newacronym{drs}{DRS}{Discovery Reference Signal}
\newacronym{du}{DU}{Distributed Unit}
\newacronym{e2e}{E2E}{end-to-end}
\newacronym{ecaas}{ECaaS}{Edge-Cloud-as-a-Service}
\newacronym{ecn}{ECN}{Explicit Congestion Notification}
\newacronym{edf}{EDF}{Earliest Deadline First}
\newacronym{embb}{eMBB}{Enhanced Mobile Broadband}
\newacronym{empower}{EMPOWER}{EMpowering transatlantic PlatfOrms for advanced WirEless Research}
\newacronym{enb}{eNB}{evolved Node Base}
\newacronym{endc}{EN-DC}{E-UTRAN-\gls{nr} \gls{dc}}
\newacronym{epc}{EPC}{Evolved Packet Core}
\newacronym{eps}{EPS}{Evolved Packet System}
\newacronym{es}{ES}{Edge Server}
\newacronym{etsi}{ETSI}{European Telecommunications Standards Institute}
\newacronym[firstplural=Estimated Times of Arrival (ETAs)]{eta}{ETA}{Estimated Time of Arrival}
\newacronym{eutran}{E-UTRAN}{Evolved Universal Terrestrial Access Network}
\newacronym{faas}{FaaS}{Function-as-a-Service}
\newacronym{fapi}{FAPI}{Functional Application Platform Interface}
\newacronym{fcc}{FCC}{Federal Communications Commission}
\newacronym{fdd}{FDD}{Frequency Division Duplexing}
\newacronym{fdm}{FDM}{Frequency Division Multiplexing}
\newacronym{fdma}{FDMA}{Frequency Division Multiple Access}
\newacronym{fed4fire}{FED4FIRE+}{Federation 4 Future Internet Research and Experimentation Plus}
\newacronym{fir}{FIR}{Finite Impulse Response}
\newacronym{fit}{FIT}{Future \acrlong{iot}}
\newacronym{fpga}{FPGA}{Field Programmable Gate Array}
\newacronym{fr2}{FR2}{Frequency Range 2}
\newacronym{fs}{FS}{Fast Switching}
\newacronym{fscc}{FSCC}{Flow Sharing Congestion Control}
\newacronym{ftp}{FTP}{File Transfer Protocol}
\newacronym{fw}{FW}{Flow Window}
\newacronym{ge}{GE}{Gaussian Elimination}
\newacronym{gnb}{gNB}{Next Generation Node Base}
\newacronym{gop}{GOP}{Group of Pictures}
\newacronym{gpr}{GPR}{Gaussian Process Regressor}
\newacronym{gpu}{GPU}{Graphics Processing Unit}
\newacronym{gtp}{GTP}{GPRS Tunneling Protocol}
\newacronym{gtpc}{GTP-C}{GPRS Tunnelling Protocol Control Plane}
\newacronym{gtpu}{GTP-U}{GPRS Tunnelling Protocol User Plane}
\newacronym{gtpv2c}{GTPv2-C}{\gls{gtp} v2 - Control}
\newacronym{gw}{GW}{Gateway}
\newacronym{harq}{HARQ}{Hybrid Automatic Repeat reQuest}
\newacronym{hetnet}{HetNet}{Heterogeneous Network}
\newacronym{hh}{HH}{Hard Handover}
\newacronym{hol}{HOL}{Head-of-Line}
\newacronym{hqf}{HQF}{Highest-quality-first}
\newacronym{hss}{HSS}{Home Subscription Server}
\newacronym{http}{HTTP}{HyperText Transfer Protocol}
\newacronym{ia}{IA}{Initial Access}
\newacronym{iab}{IAB}{Integrated Access and Backhaul}
\newacronym{ic}{IC}{Incident Command}
\newacronym{ietf}{IETF}{Internet Engineering Task Force}
\newacronym{imsi}{IMSI}{International Mobile Subscriber Identity}
\newacronym{imt}{IMT}{International Mobile Telecommunication}
\newacronym{iot}{IoT}{Internet of Things}
\newacronym{ip}{IP}{Internet Protocol}
\newacronym{itu}{ITU}{International Telecommunication Union}
\newacronym{kpi}{KPI}{Key Performance Indicator}
\newacronym{kvm}{KVM}{Kernel-based Virtual Machine}
\newacronym{los}{LOS}{Line-of-Sight}
\newacronym{lsm}{LSM}{Link-to-System Mapping}
\newacronym{lstm}{LSTM}{Long Short Term Memory}
\newacronym{lte}{LTE}{Long Term Evolution}
\newacronym{lxc}{LXC}{Linux Container}
\newacronym{m2m}{M2M}{Machine to Machine}
\newacronym{mac}{MAC}{Medium Access Control}
\newacronym{manet}{MANET}{Mobile Ad Hoc Network}
\newacronym{mano}{MANO}{Management and Orchestration}
\newacronym{mc}{MC}{Multi-Connectivity}
\newacronym{mcc}{MCC}{Mobile Cloud Computing}
\newacronym{mchem}{MCHEM}{Massive Channel Emulator}
\newacronym{mcs}{MCS}{Modulation and Coding Scheme}
\newacronym{mec}{MEC}{Multi-access Edge Computing}
\newacronym{mec2}{MEC}{Mobile Edge Cloud}
\newacronym{mfc}{MFC}{Mobile Fog Computing}
\newacronym{mi}{MI}{Mutual Information}
\newacronym{mib}{MIB}{Master Information Block}
\newacronym{miesm}{MIESM}{Mutual Information Based Effective SINR}
\newacronym{mimo}{MIMO}{Multiple Input, Multiple Output}
\newacronym{mgen}{MGEN}{Multi-Generator}
\newacronym{ml}{ML}{Machine Learning}
\newacronym{mlr}{MLR}{Maximum-local-rate}
\newacronym[plural=\gls{mme}s,firstplural=Mobility Management Entities (MMEs)]{mme}{MME}{Mobility Management Entity}
\newacronym{mmtc}{mMTC}{Massive Machine-Type Communications}
\newacronym{mmwave}{mmWave}{millimeter wave}
\newacronym{mpdccp}{MP-DCCP}{Multipath Datagram Congestion Control Protocol}
\newacronym{mptcp}{MPTCP}{Multipath TCP}
\newacronym{mr}{MR}{Maximum Rate}
\newacronym{mrdc}{MR-DC}{Multi \gls{rat} \gls{dc}}
\newacronym{mse}{MSE}{Mean Square Error}
\newacronym{mss}{MSS}{Maximum Segment Size}
\newacronym{mt}{MT}{Mobile Termination}
\newacronym{mtd}{MTD}{Machine-Type Device}
\newacronym{mtu}{MTU}{Maximum Transmission Unit}
\newacronym{mumimo}{MU-MIMO}{Multi-user \gls{mimo}}
\newacronym{mvno}{MVNO}{Mobile Virtual Network Operator}
\newacronym{nalu}{NALU}{Network Abstraction Layer Unit}
\newacronym{nas}{NAS}{Network Attached Storage}
\newacronym{nbiot}{NB-IoT}{Narrow Band IoT}
\newacronym{nfv}{NFV}{Network Function Virtualization}
\newacronym{nfvi}{NFVI}{Network Function Virtualization Infrastructure}
\newacronym{nic}{NIC}{Network Interface Card}
\newacronym{nlos}{NLOS}{Non-Line-of-Sight}
\newacronym{now}{NOW}{Non Overlapping Window}
\newacronym{nrdz}{NRDZ}{National Radio Dynamic Zone}
\newacronym{nsf}{NSF}{National Science Foundation}
\newacronym{nsm}{NSM}{Network Service Mesh}
\newacronym[type=hidden]{nr}{NR}{New Radio}
\newacronym{nrf}{NRF}{Network Repository Function}
\newacronym{nsa}{NSA}{Non Stand Alone}
\newacronym{nse}{NSE}{Network Slicing Engine}
\newacronym{nssf}{NSSF}{Network Slice Selection Function}
\newacronym{ntp}{NTP}{Network Time Protocol}
\newacronym{o2i}{O2I}{Outdoor to Indoor}
\newacronym{oai}{OAI}{OpenAirInterface}
\newacronym{oaicn}{OAI-CN}{\gls{oai} \acrlong{cn}}
\newacronym{oairan}{OAI-RAN}{\acrlong{oai} \acrlong{ran}}
\newacronym{oam}{OAM}{Operations, Administration and Maintenance}
\newacronym{ofdm}{OFDM}{Orthogonal Frequency Division Multiplexing}
\newacronym{olia}{OLIA}{Opportunistic Linked Increase Algorithm}
\newacronym{omec}{OMEC}{Open Mobile Evolved Core}
\newacronym{onap}{ONAP}{Open Network Automation Platform}
\newacronym{onf}{ONF}{Open Networking Foundation}
\newacronym{onos}{ONOS}{Open Networking Operating System}
\newacronym{oom}{OOM}{\gls{onap} Operations Manager}
\newacronym{opnfv}{OPNFV}{Open Platform for \gls{nfv}}
\newacronym[type=hidden]{oran}{O-RAN}{Open \gls{ran}}
\newacronym{orbit}{ORBIT}{Open-Access Research Testbed for Next-Generation Wireless Networks}
\newacronym{os}{OS}{Operating System}
\newacronym{oss}{OSS}{Operations Support System}
\newacronym{pa}{PA}{Position-aware}
\newacronym{pase}{PASE}{Prioritization, Arbitration, and Self-adjusting Endpoints}
\newacronym{pawr}{PAWR}{Platforms for Advanced Wireless Research}
\newacronym{pbch}{PBCH}{Physical Broadcast Channel}
\newacronym{pcef}{PCEF}{Policy and Charging Enforcement Function}
\newacronym{pcfich}{PCFICH}{Physical Control Format Indicator Channel}
\newacronym{pcrf}{PCRF}{Policy and Charging Rules Function}
\newacronym{pdcch}{PDCCH}{Physical Downlink Control Channel}
\newacronym{pdcp}{PDCP}{Packet Data Convergence Protocol}
\newacronym{pdsch}{PDSCH}{Physical Downlink Shared Channel}
\newacronym{pdu}{PDU}{Packet Data Unit}
\newacronym{pf}{PF}{Proportional Fair}
\newacronym{pgw}{PGW}{Packet Gateway}
\newacronym{phich}{PHICH}{Physical Hybrid ARQ Indicator Channel}
\newacronym{phy}{PHY}{Physical}
\newacronym{pmch}{PMCH}{Physical Multicast Channel}
\newacronym{pmi}{PMI}{Precoding Matrix Indicators}
\newacronym{powder}{POWDER}{Platform for Open Wireless Data-driven Experimental Research}
\newacronym{ppo}{PPO}{Proximal Policy Optimization}
\newacronym{ppp}{PPP}{Poisson Point Process}
\newacronym{prach}{PRACH}{Physical Random Access Channel}
\newacronym{prb}{PRB}{Physical Resource Block}
\newacronym{psnr}{PSNR}{Peak Signal to Noise Ratio}
\newacronym{pss}{PSS}{Primary Synchronization Signal}
\newacronym{pucch}{PUCCH}{Physical Uplink Control Channel}
\newacronym{pusch}{PUSCH}{Physical Uplink Shared Channel}
\newacronym{qam}{QAM}{Quadrature Amplitude Modulation}
\newacronym{qci}{QCI}{\gls{qos} Class Identifier}
\newacronym{qoe}{QoE}{Quality of Experience}
\newacronym{qos}{QoS}{Quality of Service}
\newacronym{quic}{QUIC}{Quick UDP Internet Connections}
\newacronym{rach}{RACH}{Random Access Channel}
\newacronym{ran}{RAN}{Radio Access Network}
\newacronym[firstplural=Radio Access Technologies (RATs)]{rat}{RAT}{Radio Access Technology}
\newacronym{rcn}{RCN}{Research Coordination Network}
\newacronym{rec}{REC}{Radio Edge Cloud}
\newacronym{red}{RED}{Random Early Detection}
\newacronym{renew}{RENEW}{Reconfigurable Eco-system for Next-generation End-to-end Wireless}
\newacronym{rf}{RF}{Radio Frequency}
\newacronym{rfc}{RFC}{Request for Comments}
\newacronym{rfr}{RFR}{Random Forest Regressor}
\newacronym{ric}{RIC}{\gls{ran} Intelligent Controller}
\newacronym{rlc}{RLC}{Radio Link Control}
\newacronym{rlf}{RLF}{Radio Link Failure}
\newacronym{rlnc}{RLNC}{Random Linear Network Coding}
\newacronym{rmse}{RMSE}{Root Mean Squared Error}
\newacronym{rnis}{RNIS}{Radio Network Information Service}
\newacronym{rr}{RR}{Round Robin}
\newacronym{rrc}{RRC}{Radio Resource Control}
\newacronym{rrm}{RRM}{Radio Resource Management}
\newacronym{rru}{RRU}{Remote Radio Unit}
\newacronym{rs}{RS}{Remote Server}
\newacronym{rsrp}{RSRP}{Reference Signal Received Power}
\newacronym{rsrq}{RSRQ}{Reference Signal Received Quality}
\newacronym{rss}{RSS}{Received Signal Strength}
\newacronym{rssi}{RSSI}{Received Signal Strength Indicator}
\newacronym{rtt}{RTT}{Round Trip Time}
\newacronym{ru}{RU}{Radio Unit}
\newacronym{rw}{RW}{Receive Window}
\newacronym{rx}{RX}{Receiver}
\newacronym{s1ap}{S1AP}{S1 Application Protocol}
\newacronym{sa}{SA}{standalone}
\newacronym{sack}{SACK}{Selective Acknowledgment}
\newacronym{sap}{SAP}{Service Access Point}
\newacronym{sc2}{SC2}{Spectrum Collaboration Challenge}
\newacronym{scef}{SCEF}{Service Capability Exposure Function}
\newacronym{sch}{SCH}{Secondary Cell Handover}
\newacronym{scoot}{SCOOT}{Split Cycle Offset Optimization Technique}
\newacronym{sctp}{SCTP}{Stream Control Transmission Protocol}
\newacronym{sdap}{SDAP}{Service Data Adaptation Protocol}
\newacronym{sdk}{SDK}{Software Development Kit}
\newacronym{sdm}{SDM}{Space Division Multiplexing}
\newacronym{sdma}{SDMA}{Spatial Division Multiple Access}
\newacronym{sdn}{SDN}{Software-defined Networking}
\newacronym{sdr}{SDR}{Software-defined Radio}
\newacronym{seba}{SEBA}{SDN-Enabled Broadband Access}
\newacronym{sgsn}{SGSN}{Serving GPRS Support Node}
\newacronym{sgw}{SGW}{Service Gateway}
\newacronym{si}{SI}{Study Item}
\newacronym{sib}{SIB}{Secondary Information Block}
\newacronym{sinr}{SINR}{Signal to Interference plus Noise Ratio}
\newacronym{sip}{SIP}{Session Initiation Protocol}
\newacronym{siso}{SISO}{Single Input, Single Output}
\newacronym{sla}{SLA}{Service Level Agreement}
\newacronym{sm}{SM}{Saturation Mode}
\newacronym{smf}{SMF}{Session Management Function}
\newacronym{smo}{SMO}{Service Management and Orchestration}
\newacronym{sms}{SMS}{Short Message Service}
\newacronym{smsgmsc}{SMS-GMSC}{\gls{sms}-Gateway}
\newacronym{snr}{SNR}{Signal-to-Noise-Ratio}
\newacronym{son}{SON}{Self-Organizing Network}
\newacronym{sptcp}{SPTCP}{Single Path TCP}
\newacronym{srb}{SRB}{Service Radio Bearer}
\newacronym{srn}{SRN}{Standard Radio Node}
\newacronym{srs}{SRS}{Sounding Reference Signal}
\newacronym{ss}{SS}{Synchronization Signal}
\newacronym{sss}{SSS}{Secondary Synchronization Signal}
\newacronym{st}{ST}{Spanning Tree}
\newacronym{svc}{SVC}{Scalable Video Coding}
\newacronym{tb}{TB}{Transport Block}
\newacronym{tcp}{TCP}{Transmission Control Protocol}
\newacronym{tdd}{TDD}{Time Division Duplexing}
\newacronym{tdm}{TDM}{Time Division Multiplexing}
\newacronym{tdma}{TDMA}{Time Division Multiple Access}
\newacronym{tfl}{TfL}{Transport for London}
\newacronym{tfrc}{TFRC}{TCP-Friendly Rate Control}
\newacronym{tft}{TFT}{Traffic Flow Template}
\newacronym{tgen}{TGEN}{Traffic Generator}
\newacronym{tip}{TIP}{Telecom Infra Project}
\newacronym{tm}{TM}{Transparent Mode}
\newacronym{to}{TO}{Telco Operator}
\newacronym{tr}{TR}{Technical Report}
\newacronym{trp}{TRP}{Transmitter Receiver Pair}
\newacronym{ts}{TS}{Technical Specification}
\newacronym{tti}{TTI}{Transmission Time Interval}
\newacronym{ttt}{TTT}{Time-to-Trigger}
\newacronym{tx}{TX}{Transmitter}
\newacronym{uas}{UAS}{Unmanned Aerial System}
\newacronym{uav}{UAV}{Unmanned Aerial Vehicle}
\newacronym{udm}{UDM}{Unified Data Management}
\newacronym{udp}{UDP}{User Datagram Protocol}
\newacronym{udr}{UDR}{Unified Data Repository}
\newacronym{ue}{UE}{User Equipment}
\newacronym{uhd}{UHD}{\gls{usrp} Hardware Driver}
\newacronym{ul}{UL}{Uplink}
\newacronym{um}{UM}{Unacknowledged Mode}
\newacronym{uml}{UML}{Unified Modeling Language}
\newacronym{upa}{UPA}{Uniform Planar Array}
\newacronym{upf}{UPF}{User Plane Function}
\newacronym{urllc}{URLLC}{Ultra Reliable and Low Latency Communication}
\newacronym{usa}{U.S.}{United States}
\newacronym{usim}{USIM}{Universal Subscriber Identity Module}
\newacronym{usrp}{USRP}{Universal Software Radio Peripheral}
\newacronym{utc}{UTC}{Urban Traffic Control}
\newacronym{vim}{VIM}{Virtualization Infrastructure Manager}
\newacronym{vm}{VM}{Virtual Machine}
\newacronym{vnf}{VNF}{Virtual Network Function}
\newacronym{volte}{VoLTE}{Voice over \gls{lte}}
\newacronym{voltha}{VOLTHA}{Virtual OLT HArdware Abstraction}
\newacronym{vr}{VR}{Virtual Reality}
\newacronym{vran}{vRAN}{Virtualized \gls{ran}}
\newacronym{vss}{VSS}{Video Streaming Server}
\newacronym{wbf}{WBF}{Wired Bias Function}
\newacronym{wf}{WF}{Wired-first}
\newacronym{wlan}{WLAN}{Wireless Local Area Network}
\newacronym{osm}{OSM}{Open Source \gls{nfv} Management and Orchestration}
\newacronym{pnf}{PNF}{Physical Network Function}
\newacronym{drl}{DRL}{Deep Reinforcement Learning}
\newacronym{mtc}{MTC}{Machine-type Communications}

\newacronym{cif}{CI}{cyberinfrastructure}
\newacronym{sonic}{SONiC}{Software for Open Networking in the Cloud}
\newacronym{ocp}{OCP}{Open Compute Project}
\newacronym{snmp}{SNMP}{Simple Network Management Protocol}
\newacronym{raid}{RAID}{redundant array of independent disks}
\newacronym{nfs}{NFS}{Network File Storage}
\newacronym{ci}{CI}{Continuous Integration}
\newacronym{cd}{CD}{Continuous Deployment}
\newacronym{dtn}{DTN}{Data Transfer Node}

\newacronym{dns}{DNS}{Domain Name Service}
\newacronym{nrpe}{NRPE}{Nagios Remote Plugin Executor}
\newacronym{ldap}{LDAP}{Lightweight Directory Access Protocol}
\newacronym{lan}{LAN}{Local Area Network}
\newacronym{vlan}{VLAN}{Virtual LAN}

\newacronym{ipmi}{IPMI}{Intelligent Platform Management Interface}
\newacronym{tor}{ToR}{Top-of-the-Rack}
\newacronym{lmn}{LMN}{Large Memory Node}
\newacronym{bgp}{BGP}{Border Gateway Protocol}
\newacronym{dhcp}{DHCP}{Dynamic Host Configuration Protocol}
\newacronym{vrf}{VRF}{Virtual Routing and Forwarding}
\newacronym{vpn}{VPN}{Virtual Private Network}
\newacronym{rma}{RMA}{Return Merchandise Authorization}
\newacronym{hpc}{HPC}{High Performance Compute}

\newacronym{nu}{NU}{Northeastern University}
\newacronym{asic}{ASIC}{Application-specific Integrated Circuit}
\newacronym{rdma}{RDMA}{Remote Direct Memory Access}
\newacronym{roce}{RoCE}{RDMA over Converged Ethernet}
\newacronym{ovs}{OVS}{Open vSwitch}
\newacronym{frr}{FRR}{Free Range Routing}
\newacronym{ups}{UPS}{Uninterruptible Power Supply}

\newacronym{ntia}{NTIA}{National Telecommunications and Information Administration}
\newacronym{pii}{PII}{Personal and Identifiable Information}
\newacronym{irb}{IRB}{Institutional Review Board}
\newacronym{doi}{DOI}{Digital Object Identifier}

\newacronym{sdo}{SDO}{Standard-Development Organization}
\newacronym{ose}{OSE}{Open Source Ecosystem}
\newacronym{osc}{OSC}{O-RAN Software Community}
\newacronym{dop}{DOP}{Director of Operations}
\newacronym{pm}{PM}{Program Manager}
\newacronym{excom}{EXCOM}{Executive Committee}
\newacronym{iiot}{IIoT}{Industrial \gls{iot}}
\newacronym{lf}{LF}{Linux Foundation}

\newacronym{wiot}{WIoT}{Institute for the Wireless Internet of Things}

\newacronym{otic}{OTIC}{Open Testing \& Integration Centre}

\newacronym{nofo}{NOFO}{Notice of Funding Opportunity}

\newacronym{onr}{ONR}{Office of Naval Research}
\newacronym{afosr}{AFOSR}{Air Force Office of Scientific Research}
\newacronym{afrl}{AFRL}{Air Force Research Laboratory}
\newacronym{arl}{ARL}{Army Research Laboratory}

\newacronym{arc}{ARC}{Aerial Research Cloud}

\newacronym{mno}{MNO}{Mobile Network Operator}
\newacronym{ct}{CT}{Continuous Testing}
\newacronym{oci}{OCI}{Open Container Initiative}
\newacronym{macsec}{MACsec}{Media Access Control Security}
\newacronym{pt}{PT}{Plain Text}
\newacronym{cuda}{CUDA}{Compute Unified Device Architecture}
\newacronym{cbrs}{CBRS}{Citizen Broadband Radio Service}
\newacronym{sas}{SAS}{Spectrum Access System}
\newacronym{rfi}{RFI}{Radio-Frequency Interference}
\newacronym{pal}{PAL}{Priority Access License}
\newacronym{gaa}{GAA}{General Authorized Access}
\newacronym{esc}{ESC}{Environmental Sensing Capability}
\newacronym{ota}{OTA}{Over-the-Air}

\IEEEoverridecommandlockouts
\usepackage{bm,cite}
\usepackage{amsmath}
\usepackage{amsbsy}
\usepackage{latexsym}
\usepackage{amssymb}
\usepackage{wasysym}
\usepackage{mathtools}

\DeclareMathAlphabet{\mathbit}{OML}{cmr}{bx}{it}
\DeclareMathAlphabet{\mathsf}{OT1}{cmss}{m}{n}
\DeclareMathAlphabet{\mathTXf}{OT1}{cmss}{bx}{it}

\DeclareMathAlphabet{\mathpzc}{OT1}{pzc}{m}{it}

\title{Listen-While-Talking: Toward dApp-based Real-Time Spectrum Sharing in O-RAN\vspace{-.3cm}}

\author{
\IEEEauthorblockN{Rajeev Gangula\IEEEauthorrefmark{1}, Andrea Lacava\IEEEauthorrefmark{1}\IEEEauthorrefmark{2}, Michele Polese\IEEEauthorrefmark{1}, Salvatore D'Oro\IEEEauthorrefmark{1},\\Leonardo Bonati\IEEEauthorrefmark{1}, Florian Kaltenberger\IEEEauthorrefmark{1}, Pedram Johari\IEEEauthorrefmark{1}, Tommaso Melodia\IEEEauthorrefmark{1}}
\IEEEauthorblockA{\IEEEauthorrefmark{1}Institute for the Wireless Internet of Things, Northeastern University, Boston, USA, \IEEEauthorrefmark{2}Sapienza University of Rome, Italy\\
}
\thanks{This work was partially supported by OUSD(R\&E) through Army Research Laboratory Cooperative Agreement Number W911NF-24-2-0065. The views and conclusions contained in this document are those of the authors and should not be interpreted as representing the official policies, either expressed or implied, of the Army Research Laboratory or the U.S. Government. The U.S. Government is authorized to reproduce and distribute reprints for Government purposes notwithstanding any copyright notation herein.}
}

\usepackage{tikzpagenodes,etoolbox}
\usetikzlibrary{calc}
\usepackage[contents={}]{background}
\AddEverypageHook{%
\ifnumequal{\thepage}{1}{%
    \tikz[remember picture,overlay]{%
        \node[draw,
        minimum width=1.03\textwidth,
        text width=1.02\textwidth,
        font=\scriptsize
        ]
        at ($(current page header area) - (0,5pt)$)
        {%
        This work has been submitted to the IEEE for possible publication.\\
        Copyright may be transferred without notice, after which this version may no longer be accessible.
        };
    }%
}{}%end ifnumequal
}

\begin{document}

\makeatletter
\patchcmd{\@maketitle}
  {\addvspace{0.5\baselineskip}\egroup} %0.5
  {\addvspace{-1.51\baselineskip}\egroup} %-1.5
  {}
  {}
\makeatother

\maketitle

\begin{abstract}
This demo paper presents a dApp-based real-time spectrum sharing scenario where a \gls{5g} base station implementing the NR stack adapts its transmission and reception strategies based on the incumbent priority users in the \gls{cbrs} band. The dApp is responsible for obtaining relevant measurements from the \gls{gnb}, running the spectrum sensing inference, and configuring the \gls{gnb} with a control action upon detecting the primary incumbent user transmissions.
This approach is built on dApps, which extend the O-RAN framework to the real-time and user plane domains. Thus, it avoids the need of dedicated \glspl{sas} in the \gls{cbrs} band. The demonstration setup is based on the open-source 5G \gls{oai} framework, where we have implemented a dApp interfaced with a \gls{gnb} and communicating with a \gls{cots} \gls{ue} in an over-the-air wireless environment.
When an incumbent user has active transmission, the dApp will detect and inform the primary user presence to the \gls{gnb}. The dApps will also enforce a control policy that adapts the scheduling and transmission policy of the \gls{ran}.
This demo provides valuable insights into the potential of using dApp-based spectrum sensing with O-RAN architecture in next generation cellular networks.
\end{abstract} 
%%--------------------------------------------------------------------%%

\glsresetall

\begin{figure*}[h]
    \centering  
    \includegraphics[width=0.9\textwidth]{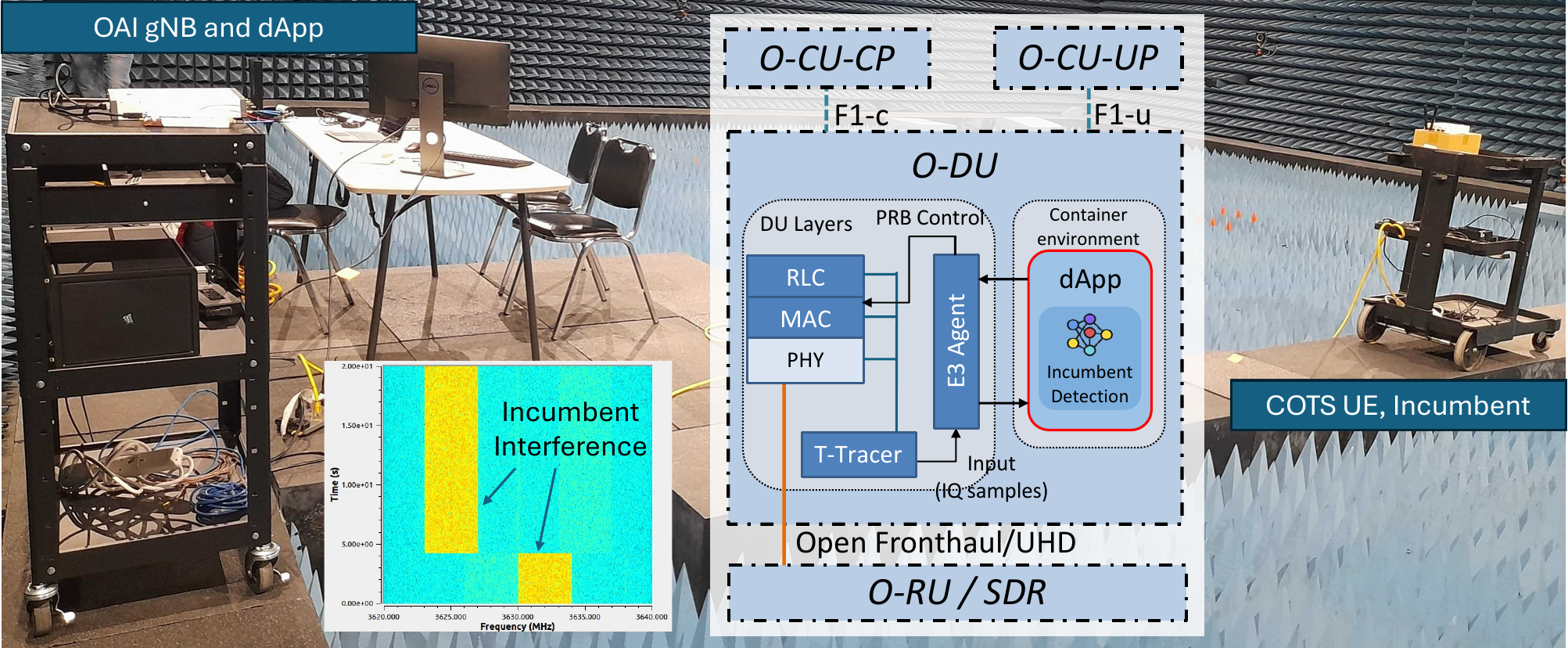}
    \setlength\belowcaptionskip{-.5cm}
    \setlength\abovecaptionskip{0.1cm}
    \caption{dApp-based spectrum sharing architecture and \gls{ota} demo.}
    \label{fig:sys_mod}
\end{figure*}

% \begin{figure*}[h]
%     \centering  
%     \includegraphics[width=.4\textwidth]{IMG/dapp_arch.png}
%     \caption{dApp-based spectrum sharing architecture}
%     \label{fig:sys_mod}
% \end{figure*}

\section{Introduction} \label{sec:intro}

The radio spectrum is a valuable and
finite resource. The proliferation of systems and wireless technologies supporting various users, devices, and machines in this era of connected intelligence has led to spectrum congestion, especially in the frequency bands below 6 GHz. The traditional approach of dividing the spectrum into licensed and unlicensed bands has resulted in under-utilization as it is unable to exploit the spatial and temporal variations in the spectrum demands. To this end, dynamic spectrum sharing has emerged as a key technology and it is expected to become
a fundamental part of next-generation wireless system design.

In a spectrum sharing scenario, multiple categories of users, likely leveraging different wireless access technologies,
dynamically select the unused portion of the spectrum while satisfying the sharing constraints such as access priorities and interference limits.
A popular spectrum sharing example in the U.S
is the \gls{cbrs}, in the 3.55-3.7 GHz frequency range. Mobile network operators (also known as \gls{pal} users) and unlicensed users, or \gls{gaa} users, may coexist with high priority incumbent federal radar and satellite systems users. When considering 5G NR systems, the \gls{cbrs} band is a subset of the bands n77 and n78.

Spectrum sharing in \gls{cbrs} is managed by a centralized \gls{sas}, with the aid of dedicated \gls{esc} sensor networks \cite{ColTriSilva_23}.
The \gls{sas} requires \gls{pal} and
\gls{gaa} users to register to a database and coordinate the spectrum access based on the priority rules. The \gls{esc} network is used for spectrum sensing, to detect the primary user presence in case this is not advertised beforehand.
The drawback of this method is the need of a dedicated sensing infrastructure and coordination of various entities belonging to different wireless access technologies by a central station. In addition, \gls{cbrs} systems operate in a timescale of minutes, not allowing for the subframe-level sharing that would improve spectrum utilization through statistical multiplexing.

On the other hand, the Open \gls{ran} paradigm, implemented through the specifications of the O-RAN ALLIANCE, has introduced the possibility of performing closed-loop control and adaption of the \gls{ran}, as well as open, virtualized, disaggregated wireless networks. The intelligent control loops have been shown as enablers of several dynamic optimization use cases, toward a data-driven \gls{ran} with bespoke configurations and optimized performance~\cite{polese2023understanding,rob2021oran}. 
The current O-RAN architecture, however, does not extend to the user plane, wherein lies key information that can be used for spectrum sensing, and to real-time control, necessary to implement strategies that promptly adapt the \gls{gnb} stack to avoid interference.
% With the shared cellular infrastructure and intelligent control loops, O-RAN based spectrum sensing and coordination can eliminate the drawback of existing SAS method.
% The base stations (gNBs) in a 5G network can potentially be used for communication and sensing \cite{RenUpUtNatNasShaChow_23}. 

In this work, we build on the concept of dApps, a real-time and user-plane extension of the O-RAN architecture we proposed in~\cite{DorPolBonHaMel_22}, to introduce the ``listen-while-talking'' approach for real-time \gls{ran}-driven spectrum sharing. With this, \glspl{gnb} can communicate while simultaneously performing the spectrum sensing task in the \gls{cbrs} band, effectively configuring the whole \gls{ran} as a spectrum sensor. dApps, which are currently studied as components for the next-generation of O-RAN systems, interface with the gNB stack to (i) extract I/Q samples from dedicated symbols reserved for spectrum sensing at the \gls{gnb}; (ii) perform data-driven inference on whether the spectrum is available, or, in case an incumbent is detected, to determine which portion of the spectrum needs to be vacated; and (iii) to inform the \gls{gnb} about possible control actions that need to be adopted to avoid the incumbent. Here, we consider barring the scheduling over the \glspl{prb} where an incumbent user is detected. This allows for the minimum disruption of cellular services while vacating the spectrum for the incumbent user. 
% Upon detecting an incumbent user, the gNB alters its scheduling policies and parameters such that cellular connectivity is maintained with minimum disruption while vacating the spectrum for the incumbent user. 
% We demonstrate a spectrum sensing distributed applications (dApp) \cite{DorPolBonHaMel_22} that
% extracts the IQ samples from the dedicated symbols reserved for spectrum sensing at the gNB, performs the inference, and informs the gNB about the barred physical resource blocks (PRBs) where an incumbent user is detected. 
To the best of our knowledge, an O-RAN-based real-time spectrum sharing system with a 5G \gls{gnb} operating in the \gls{cbrs} band has not been demonstrated before.

\section{System Architecture} \label{sec:sysarch}
The proposed spectrum sharing framework is illustrated in Figure~\ref{fig:sys_mod}. It leverages the following components.
% The individual components are described in the following sections.

% \subsection{5G NR}
\textbf{Programmable 5G RAN.}
The 5G network comprises of a core network and \gls{gnb}
running in the standalone mode, using the open-source \gls{oai} stack. The \gls{gnb} uses the \gls{tdd} mode and is configured with \textit{sensing symbols} in some of the \gls{ul} slots.
These are dedicated for spectrum sensing, and the gNB does not schedule any \glspl{ue} in these symbols, for any of its uplink channels. 
% The gNB is configured such that there are no UL physical channels PUSCH, PUCCH and SRS is configured in these symbols.
In this demo, we configure one \gls{ofdm} symbol per frame to be used for sensing. This leads to periodic spectrum sensing over the entire gNB bandwidth every 10 ms, i.e., an overhead of 0.35\%, or 1 symbol every 280 with NR numerology 1.
%For example if the gNB is configured with a TDD pattern , i.e., an overhead of \hl{XYZ}.
Note that this configuration is programmable. 
Once the I/Q samples in the \textit{sensing symbols} are available, they are shared with a dApp, where the inference is performed.
% The frequency domain IQ samples extracted from the \textit{sensing symbols} are
% collected by a dApp where the spectrum sensing inference is performed. 
% The detailed description of the dApps is given in the next section. 
To this purpose, we implemented an \textit{E3 agent} in the OAI stack. 
% The OAI gNB is augmented with an E3 agent, through which it interacts with the dApps.
The E3 is an O-RAN E2-like interface 
created to extract parameters from the gNB that are not exposed to the xApps, e.g., user plane elements. The E3 agent running inside the OAI gNB leverages the {\tt T\_tracer} tool to extract the data~\cite{OaiT}.
The E3 agent follows a publish/subscribe message mechanism so that several dApps can interact with the gNB while reducing the inter-dependency. In this demo, we use the E3 agent along with the dApp to extract the IQ samples.

% and the user.
% While the gNB and the core are based on OpenAirInterface (OAI) software stack, the user is a COTS UE. USRP SDR peripheral is used as RF fronted module.
% The gNB operates in the n78 band with 80 MHz bandwidth.
% The TDD mode is used with configuration xxxx.
% The gNB is configured such that on slot 8 a symbol is completely left unscheduled for users
% and essentially used for spectrum sensing purposes. This can be achieved by the PUSCH SLIV allocation parameters cite \cite{3gpp2018nr_38_214}.
% This leads to periodic spectrum sensing over the gNB bandwidth every 1 ms.

% The gNB can change some parameters and scheduling based on the inputs from the dApp which is explained in the next.

\textbf{dApp.} These programmable elements complement
existing xApps/rApps to extract real-time data and perform inference on lower-layer functionalities.  
In our implementation, the spectrum sensing dApp connects with the gNB using the E3 interface and extracts the I/Q samples from the configured 
\textit{sensing symbols}.
An inference algorithm takes this data as input and outputs the list of \glspl{prb} where the incumbent user presence is detected, which are then sent as barred \glspl{prb} to the gNB scheduler. For this demo, we leverage an energy threshold method, with more advanced solutions employing machine learning as future work. 
% The dApp informs the list to the gNB which then adapts its configuration parameters and scheduling policy to avoid these PRBs.
%Note that the proposed mechanism works only 

\section{Demonstration}\label{sec:demo}

This demonstration consists of an OAI gNB, a \gls{cots} UE and an incumbent user.
A computer running Ubuntu 22 OS attached with USRP SDR hosts the core, gNB (with E3 agent), and dApp application.
The COTS UE is connected to the gNB over-the-air and runs an {\tt iperf} session all the time. The incumbent user is modeled using a GNU-radio script hosted on a mini computer attached to a USRP. In the live experiment, we will show a) the incumbent user spectrum with dynamic changes over time; b)
spectrum sensing through the dApp; and c) changes in the UE throughput due to the barred PRBs that are occupied by the incumbent user. This demo can be run without the need of specific experimental spectrum licenses by placing the \gls{cots} UE and the antennas in a shielded RF box.

% \subsection{Experiment}\label{sec:exp}
% \begin{itemize}
%     \item Start the core network, gNB in band n78 with 80 MHz bandwidth in PC1
%     \item dApp is launched on PC1 and it interacts with the E3 agent inside the gNB
%     \item The COTS UE is connected to the gNB over-the-air
%     \item Execute the GNU radio primary user script in PC2
%     \item The dApp's incumbent user detection is plotted on screen
%     \item The UE throughput changes with an iperf application is shown in the screen
% \end{itemize}

% \section{Acknowledgement} \label{sec:ack}
% This work is supported by the OUSD(R\&E) through Army Research Laboratory Cooperative Agreement Number W911NF-19-2-0221. The views and conclusions contained in this document are those of the authors and should not be interpreted as representing the official policies, either expressed or implied, of the Army Research Laboratory or the U.S. Government.

\bibliographystyle{IEEEtran}
\bibliography{biblio}

\end{document}